
  \magnification=\magstep1
\settabs 18 \columns
 \hsize=16truecm
\input epsf

 \def\b{\bigskip}
 
\def\no{\noindent}
\def\r{\rightline}
\def\ce{\centerline}
\def\ve{\vfill\eject}

\def\today{August 10, 2005}

\r \today
 
\def\Rit{\hbox{\it I\hskip -2pt R}}

 \font\steptwo=cmb10 scaled\magstep2
\font\stepthree=cmb10 scaled\magstep4
\magnification=\magstep1
  
\vskip1in 
  {\ce {\stepthree   Growth of a Black Hole}}  
\b
 \ce{Christian Fr\o nsdal}
\b
 \ce{\it Physics Department, University of California, Los Angeles CA
 90095-1547 USA}
\b
 
 \b\b

\no{\it ABSTRACT.} This paper studies the
interpretation of physics near a Schwarzschild black hole. A scenario for
  creation and growth is proposed that avoids the conundrum of
information loss.  In this picture the horizon recedes as it is
approached and has no physical reality. Radiation is likely to occur, but
it cannot be predicted.

\b\b\b

\ce{   \bf 1. INTRODUCTION} 

Creation of a black hole - a spacetime with the Schwartzschild line
element- is a contradiction in terms. We think that it may be possible,
nevertheless to give sense to the notion, and we shall attempt to do so,
mainly by introducing an observer, in the established tradition of General
Relativity. Here is a rough summary of our proposal.

The sense of ``creation'', as ``becoming in time'' is here attributed to
an observer who is aware of a limited region of space. Over a period of
time he notes that the metric in this region is Minkowski. Then a
spherical shower arrives  from (spatial) ``infinity''.  In a moment the
shower passes the observer, and now he finds himself in a portion of
space time with a Schwarzschild metric. The further passage of time does
nothing to change this metric, except to enlarge the domain on which it
applies, unless other showers arrive, each one leading to an increase in
the mass parameter of the Schwarzschild metric near our observer.

We consider the idealized case of an
isotropic, spherical mass shell of negligible thickness. In contrast to
the arrival of a single particle, this event preserves spherical symmetry.
The description is also much simpler than the usual treatment of a
continuous distribution of in-falling matter. The metric is 
Schwarzschild (or Minkowski) inside the shell and Schwarzschild   
outside the shell as well, but with a different mass parameter.
Everything is determined by the trajectory, a function
$f$ that determines the radius of the shell as a function of time, $r =
f(t)$. The Einstein tensor is calculated and we try, without much
success,  to understand the nature of the  matter shell by inspection of
this tensor, interpreted as the energy momentum tensor of the matter
distribution.
 
In analogy with the famous result on the motion of particles, by
Einstein, Infeld and de Hoffman and others, see [EI], one might expect
that the trajectory is determined by the field equations. The function
$1/|\vec x - \vec x_0(t)|$ associated with a particle  is here replaced
by the 
 Heavyside function $\theta(r-f(t))$; both lead to the appearance of
delta functions in the field equations. We  investigate the
Bianchi identities, to conclude that no
constraints on the trajectory is implied by the field equations. The
calculations are not reported, but the conclusions
  will be supported by intuitive physical arguments.

The conclusion is that the probable trajectory of the incoming spherical
shell cannot be predicted from first principles. However, the simple
criterion that consists of limiting the velocity of propagation to the
velocity of light in the local metric strongly suggests that the
infalling matter can never catch up with the horizon of its own making.
  The fear of Hawking [H] and others, that the
matter shell may penetrate its own `shadow' is therefore unfounded.

It is perhaps interesting to try to strengthen the argument by pointing
out a simple generalization of the scenario. Imagine a spherical
distribution of matter converging towards a point. Instead of trying to
solve all the equations, for matter and for the metric, that govern this
complicated situation, suppose that matter is concentrated on a large
number of concentric spheres. Eventually a continuous distribution of
matter may be obtained by passing to a limit. Then between sphere number
$n$ and sphere number
$n+1$ the metric is sensibly assumed to be of the Schwarzschild type,
with a parameter $m_n$ and associated ``horizon" $r_n = 2m_n$. Let
$R_n(t)$ denote the radial coordinate of the $n$'th sphere at time $t$.
We do not think that it is fruitful to speculate on the equations of
motion of each matter sphere, for this would depend on its internal
constitution and subject to wide variation. Granted only that, as long as
the density is low, the atraction towards the center, caused by the
matter that lies inside, is a dominant factor.  Let us admit initial
conditions when the density is low and
$R_n(t) >> r_n$ and let us suppose that the motion is uniformly inward and
radial. It is conceivable that a shell may overtake another shell. It is
also possible that shell $n$ may pass through the horizon of shell $n'$.
Remember that the horizon of shell $n'$ is a purely fictitious notion
that appears in the analytic continuation of a metric that is physical
only in the interval between spheres $n'$ and $n'+1$. It would acquire
physical reality only in the event that shell $n'$ caches up with its own
horizon, and that we have argued to be very unlikely.

What will happen `eventually'? Well certainly the most likely
possibility is that the inward motion cease, due to the increasing
pressure or by quantum effects. But even if it continues indefinitely
there is no difficulty; if the radial velocity does not change sign then
it certainly tends to zero, since it takes an infinite time for even a
light signal to reach a true horizon.

It might be interesting to extend these considerations to the case of the
Kerr metric, for in that case an analysis along the lines of the Membrane
Paradigm would give additional insight. We do not expect, however, that
a probability-conserving quantum mechanics exists in the Kerr metric.   
 
 \ve

\ce{  \bf 2. THE ETERNAL BLACK HOLE} 

\no{\it 2.1. Completion of the metric.} 

Schwarzschild  
found the first nontrivial solution of Einstein's equations in vacuum
[S]. It is static and spherically symmetric and was given by
Schwarzshild in the form
$$
ds^2 = (1-2m/r)dt^2 - (1-2m/r)^{-1}dr^2 - r^2d\Omega^2,
$$
with $~~ d\Omega^2 =d\theta^2 + \sin^2\theta d\varphi^2$, for the space
time
$$
-\infty<t<\infty,~~ 2m < r < \infty,~~ \Omega \in S_2.
$$   
This manifold is not geodesically complete; there are timelike geodesics
that reach the horizon at $r = 2m$ at finite proper time though always at
infinite values of the coordinate time. The restriction $r > 2m$ was
therefore received with scepticism.

Eddington [E], and Finkelstein [F], showed that the singularity at
$r = 2m$ could be removed by a change of coordinates. The new 
time coordinate $\tau$, related to Schwarzschild's time $t$ by the
formula
$$
\tau := t - \ln(r-2m),
$$
leads to the line element  
$$
ds^2_{EF} = (1-2m/r)d\tau^2 + 2r^{-1}d\tau dr - (1+2m/r)dr^2 -
r^2d\Omega^2.
$$
However, this metric  fails to be time-symmetric and future
directed timelike geodesics run inwards only. (This space really is a
black hole!) In fact, the Eddington-Finkelstein coordinates cover only
half of a more interesting, time-symmetric space time.

To visualize a curved space it is useful to imbed it in a flat space of
higher dimension. A 2-dimensional manifold can be imbedded in 3
dimensions, a 4-dimensional one in 10 dimensions. But the Schwarzschild
metric is special, it can be imbedded in a 6-dimensional Lorentzian space
time with signature (5,1). A partial imbedding was found by Kasner [K],
the complete imbedding by Fronsdal [Fr]. The complete imbedding is given
by
$$
ds^2 = dZ_1^2 - dZ_2^2 - dZ_3^2 - dZ_4^2 - dZ_5^2 - dZ_6^2,
$$
$$
Z_1 = 2\sqrt{1-1/r}\sinh{t\over 2},~~ Z_2 = 2\sqrt{1-1/r}\cosh{t\over
2},~~ Z_3 = g(r),
$$
$$
 Z_4 = r\sin\theta\sin\varphi,~
Z_5 = r\sin\theta\cos\varphi,~ Z_6 = r\cos\theta. 
$$
where $g$ is the function defined by $(dg/dr)^2 = (r^2 + r + 1)/r^3$ and
the unit of length is $2m$.

The extended space time is the surface in 6 dimensions defined 
parametrically by the equations
$$
Z_2^2 - Z_1^2 = 4(1-1/r),~~ Z_3 = g(r),~~ Z_4^2 + Z_5^2 + Z_6^2 =
r^2.\eqno(2.1)
$$
It is time symmetric and it is geometrically complete.
It can be visualized by fixing the angles so that the metric reduces to
$$
ds^2 = dZ_1^2 - dZ_2^2 - dZ_3^2
$$
on the surface
$$
Z_2^2 - Z_1^2 = 4(1-1/r),~~ Z_3 = g(r).
$$ 
The following illustrations, Fig.s 1 and 2, are taken from [Fr]. Values
of $r$ in units of $2m$.
\vskip1.1in
\def\picture #1 by #2 (#3){
  \vbox to #2{
    \hrule width #1 height 0pt depth 0pt
    \vfill
    \special{picture #3} 
    }
  }
\def\scaledpicture #1 by #2 (#3 scaled #4){{
  \dimen0=#1 \dimen1=#2
  \divide\dimen0 by 1000 \multiply\dimen0 by #4
  \divide\dimen1 by 1000 \multiply\dimen1 by #4
  \picture \dimen0 by \dimen1 (#3 scaled #4)}
  }

\parindent=1pc

\vskip-3cm
\epsfxsize.8\hsize
\centerline{\epsfbox{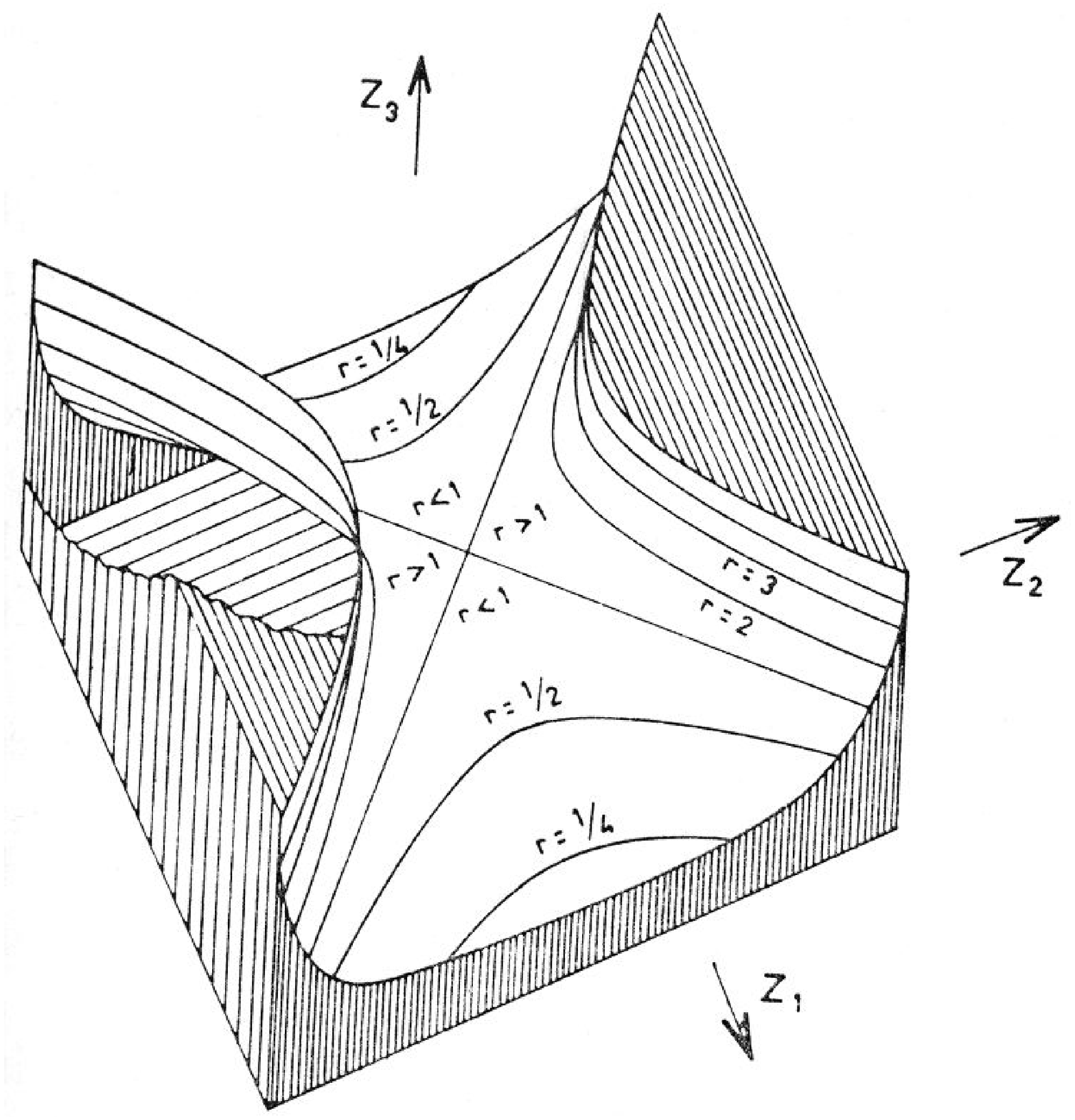}}
\vskip-1cm
\no{\it Fig 1. The surface defined by Eq.s (2.1), showing a subspace
$d\phi = d\theta = 0$ of the completed Schwarzschild manifold as a
2-dimensional surface in a pseudo-Euclidean space.}    


\parindent=1pc

\vglue2cm 
\epsfxsize.8\hsize
\centerline{\epsfbox{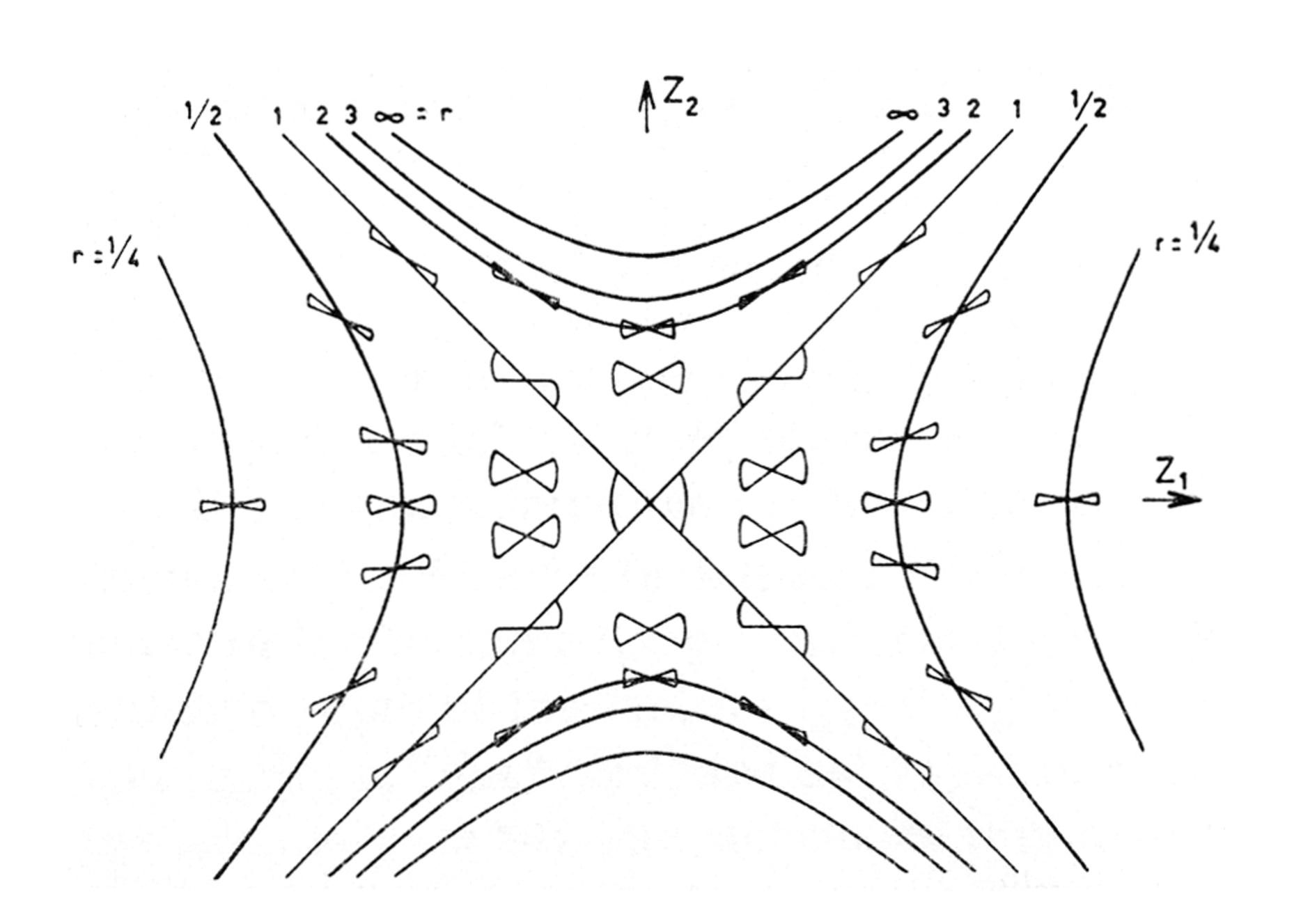}}
\no{\it Fig 2. The projection of the surface in Fig 1 on the
$Z_1,Z_2$-plane.}

 The complete manifold contains a pair of time-conjugate
Eddington-Finkelstein  manifolds. There are 4 distinct regions; two
physically equivalent outside regions and two similar inside regions, all
of them touching at the intersection of two horizons. See Fig 2. 

Timelike, future
directed geodesics  originating in the West inside   exit into the South 
outside (= physical spacetime). Some remain there, moving outwards
towards infinite distances, others enter the East inside. And some future
directed timelike paths come in from infinite distances to either go out
again or else plunge into the East inside. (We do not refer to the two
insides as `past' and `future', since the entire past and the entire
future as we can know them or study them are included in space time (the
Southern outside). 
 The appellation `black hole' is no longer apt, for the star is as much
white as it is black.

The completion was also found, by other means, by Szekeres [S] and by
Kruskal [Kr].

\b\b
\no{\it 2.2. Information and probability in an eternal black hole.}

We continue, in this Section 2, to consider the Scwarzschild solution as
a given, fixed metric, and the physics of test particles that move in it,
without taking any account of the reaction of those particles on the
gravitational metric.

A much debated question is the fate of a space traveller who penetrates
the horizon, something that seems to be allowed since there are timelike
paths that do so in finite proper time. 

That is; if we think about an observer in a space ship, for whom the
passage of time is measured by the proper time interval $ds$, then it
seems that we must accept that his fate may be to penetrate the horizon
and to remain forever inside. The trouble with this idea is that it is
not  within our rights or duties as scientists to inquire about things
that are beyond observation. Imagine a king  who sends a more modern
Columbus to investigate beyond the horizon. The intrepid traveller
dutifully sends back reports of his progress, but these reports
arrive back in court with increasing delays. To the king, fixed at home
and living in coordinate time, the reports will keep coming,\footnote*{
Unless the reports are sent with increasing frequency, traveller time,
only a finite number of reports will be received.} but always more
delayed, and always reporting from this side of the horizon. No
information from `the beyond' will ever reach the observer that remains
outside, the travellers adventures in afterlife are beyond human ken and
outside the domain with which science is rightly concerned.

Focusing our attention, as we must, on events taking place outside the
horizon, it is perfectly possible to describe such events in
terms of the coordinates; more precisely, in terms of $r,t$ and the
angles. The completion of the metric serves, not so much to discover the
existence of two outsides and two insides, but rather to dramatize the
fact that there are two horizons, a future horizon approached by ingoing
timelike paths, and a past horizon  with outcoming timelike paths. If we
limit ourselves to particle-like physical systems, and to their motion
along timelike paths, then there is no obstacle to preserving a
deterministic physical theory, no loss of information need be feared (or
invented). However, to reinforce this conviction we turn to quantum
mechanics, for it is a curious fact that quantum mechanics has been
subjected to a more extensive discussion of  interpretation and
information than classical mechanics.

Let us consider the simplest quantum mechanical system on a spacetime
with the Schwarzschild metric, keeping an open mind for a moment about
the limitation $r > 2m$. The quantum mechanical version of a theory of
point particles moving on geodesics is the theory of a scalar field $\phi$
satisfying the covariant Klein Gordon equation,
$$
({1\over \sqrt{-g}}\partial_\mu \sqrt{-g}g^{\mu\nu}\partial_\nu +
M^2)\phi(t,\vec x) = 0.
$$  
With  $\phi$ complex, this comes
from the action
$$
A = \int \sqrt{-g}(g^{\mu\nu}\partial_\mu \phi^*\partial_\nu\phi -
M^2\phi^*\phi)d^4x.
$$
The probability interpretation of quantum mechanics demands a time
independent probability density, and in this case it is given by
$$
| \phi|^2 := -i\int g^{00}(\dot \phi^* \phi - \phi^*
\dot\phi)r^2drd\Omega.
$$
The probablity density must be positive, hence $g^{00} =
1-2m/r > 0$ and $r > 2m$.

Once more we come to the inescapable conclusion that physics must concern
itself exclusively with the outside. There is great body of work by
Chandrasekhar that develops this theory of a scalar field on the outside
of the Scwarzschild horizon [C]. It is instructive to recall some of the
main features of Chandrasekhar's theory. First of all, he verifies the
time independence of the inner product,
$$
\partial_t|\phi|^2 = -i\int \Big(\partial_t(g^{00}   \phi^*) \phi -
\phi^* \partial_t (g^{00}\phi)
\Big)r^2drd\Omega.
$$
For a solution of the wave equation this becomes
$$
\partial_t|\phi|^2 = -i\int \Big(\phi^*H\phi -
(H\phi^*)\phi\Big)r^2drd\Omega,
$$ 
where $H$ is the `Hamiltonian' operator
$$
H = {1\over \sqrt{-g}}\partial_i\sqrt{-g}g^{ij}\partial_j + M^2.
$$
Positivity of the probability thus reduces to the condition that this
operator must be self adjoint. This fixes the domain of the Hamiltonian,
a space of functions that satisfy certain boundary conditions,
precisely the condition that $\phi(t,\vec x)$ vanish on the horizon. The
simplest way to demonstrate this  is to change variables,
replacing the radial coordinate $t$ by the tortoise coordinate of Regge
and Wheeler,
$$
r^* := r+ 2m\ln({r\over 2m}-1).
$$
This variable runs from $-\infty (r = 2m)$ to $+\infty (r = \infty)$. 
The inner product takes the familiar form $\int |\psi|^2 dr^* d\Omega$,
and the equation of motion for a stationary state becomes
$$
\Big({d^2\over d{r^*}^2}  - V +E^2\Big)\psi(r^*,\Omega) = 0.   
$$
with a potential $V$ that falls off as $r^* \rightarrow \pm\infty$.

This theory is characterized by a unitary S-matrix and 
the conservation of probability; there is no loss of
information. Incoming waves come from infinity 
  ($r^*
\rightarrow \infty$) \underbar{and} from the horizon ($r^*
\rightarrow -\infty)$, and
outgoing waves also approach
$r^* =
\pm
\infty$,  as in a one-dimensional scattering problem. This is in full
accord with the classical picture.

It needs to be emphasized that this whole discussion, and especially that
which follows, is placed in the context of the eternal black hole.
 
In the ordinary setting of contemporary physics one admits the existence
of cosmic rays, of particles and radiation arriving in our midst 
from distant parts of the universe. It is a useful idealization, in a
local context, to consider that cosmic rays   
originate in the infinite past at infinite distances. There is also
the infinite future, populated by the final states of the S-matrix.
In a black hole we must admit the existence of an alternative final
destiny, the future horizon, the final resting place of particles that do
not have enough energy to escape. There is also
another possible source of cosmic rays, coming out of the past horizon.
We submit that we are no more able to predict this form of radiation than
we are able to predict the ordinary cosmic rays.

Although it is probably impossible to predict the arrival of cosmic
rays, we can form a scenario that would explain their existence, in the
context of a theory of the evolution and even of the origin of the
universe. Making any prediction concerning the
radiation coming out of the past horizon of a black hole requires
a wider context, one that includes the creation of a black hole. This
concept of creation of a complete space time metric is a contradiction in
terms, nevertheless we shall try to make sense of it.

\b\b\b

\ce{   \bf 3. GROWTH OF A BLACK HOLE} 
\b
\no{\it 3.1. The metric and the Einstein tensor.}

The proposal advanced here is closely related to the ideas known under
the label Membrane Paradigm. [MPT] However, it will be convenient to
postpone   comments on this relationship till later. 

Let
$$
ds(2m)^2 := (1-2m/r) dt^2 - (1-2m/r)^{-1}dt^2 - r^2d\Omega^2,\eqno(3.1)
$$
the Schwartzschild line element with mass $m$, on the space time
$$
-\infty < t < \infty,~~ 2m<r<\infty,~~ \Omega \in S_2.
$$ 
Tolman [To] and others investigated the more general metric defined by
$$
ds^2_{AB} := {\rm e}^\nu dt^2 - {\rm e}^\lambda dr^2 -
r^2d\Omega^2,\eqno(3.2)
$$
where $\nu$ and $\lambda$ are arbitrary functions of $t$ and $r$, and
calculated the associated Einstein tensor $G_\mu^\nu = R_\mu^\nu - {1\over
2}R\delta_\mu^\nu$  (here $x^1,...,x^4 = r,\theta,\phi,t$):
$$\eqalign{&
{G_r}^r =  {\rm e}^{-\lambda }\Big({\nu'\over r} + {1\over r^2}\Big)-
{1\over r^2},\cr
& {G_t}^t =  e^{-\lambda}\Big({-\lambda'\over r}+ {1\over r^2}\Big) -
{1\over r^2},\cr
 & {G_\theta}^\theta = {G_\phi}^\phi = {\rm e}^{-\lambda}\Big({\nu''\over
2} + {\nu'^2\over 4} - {\nu'\lambda' \over 4} +{\nu'-\lambda'\over
2r}\Big) - {\rm e}^{-\nu}\Big( {\ddot\lambda\over 2} + {\dot\lambda^2\over
4}-{\dot\lambda\dot\nu\over 4}\Big),\cr &  {G_t}^r =   {\rm
e}^{-\lambda}{\dot
\lambda\over r},~~ {G_r}^t = -{\rm e}^{-\nu} {\dot\lambda\over r}.
\cr}\eqno(3.3)
$$
By Einstein's field equations this equals $-8\pi$ times the
energy-momentum tensor of all matter present.

The proposed model of mass accretion is based on the following special
case of (3.2). Let $0<r_0<r_1$ and consider the line element
$$
ds^2|_f = \Big\{\matrix{ds^2(r_0),~~~ r_0<r<f(t),\cr
\hskip-8.5mm ds^2(r_1),~~~ f(t)<r,}\eqno(3.4) 
$$
with $f: \Rit\rightarrow \Rit,~ t\mapsto f(t)$ a smooth function such that
$$
\forall t\in \Rit,~~r_1 < f(t) < \infty.
$$
In this case $\lambda = -\nu$   and (3.3) reduces to 
\footnote*{  It is remarkable that the following expressions
continue to make sense as distributions although the components of the
metric tensor are discontinuous.}
$$\eqalign{&
{G_r}^r = {G_t^t} =  (e^\nu)'/r + e^\nu/r^2-1/r^2,\cr
 & {G_\theta}^\theta = {G_\phi}^\phi 
= (e^\nu)''/2 + (e^\nu)'/r -\partial_t e^{-\nu}/2,\cr &  {G_t}^r =  
  \partial_t(e^\nu) /r,~~ {G_r}^t =  -\partial_t(e^{-\nu})/r.
\cr}\eqno(3.3)
$$
The   Einstein tensor vanishes  where $r \neq f(t)$, so
that matter is present on the moving sphere $r = f(t)$ only. Any and all
attributes of this matter that affect the metric are contained in the
trajectory function $f$.

An observer outside the matter sphere experiences a Schwartzschild metric
with mass parameter $2m_1 = r_1$, which would imply a horizon at $r =
r_1$. But to approach this apparent horizon he must first penetrate the
matter sphere at $r = f(t) > r_1$, upon which he finds  himself in a
region of space time endowed with another Scwarzschild metric; the
horizon having receeded to $r = r_0 < r_1$. In the scenario invisaged here
there is no geodesic that connects the outside world to the horizon
that appears only in the continuation of the outside metric beyond 
its region of validity. One has a Schwarzschild metric but the horizon is
ephemeral and the epithet `Black Hole' is not really appropriate.

The energy momentum tensor is
$$\eqalign{&
(T_t)^t = (T_r)^r = {r_1-r_0\over 8\pi r^2}\delta(r-f(t)),\cr
&(T_\theta)^\theta = (T_\phi)^\phi = {r_1-r_2\over 16\pi r}(1+\dot
f^2/A)\delta'(r-f(t)) - {r_1-r_0\over 16\pi Ar}\ddot f(\delta(r-f(t)),\cr
&
(T_t)^r = {r_1-r_0\over 8\pi r^2}\dot f\delta(r-f(t)),\cr
&(T_r)^t = -{r_1-r_0\over 8\pi Ar^2}\dot f \delta(r-f(t)),~~ A :=
(1-r_0/r)(1-r_1/r). 
\cr}
$$
Notice that the total energy contained in the mass shell is $(r_1-r_2)/2
= m_1-m_0$, agreeing with the increase in the Scharzschild mass parameter.

\b
\no{\it 3.2. The trajectory.}

We do not possess a convincing theory of a classical matter distribution
interacting with the gravitational field. Representing matter by a
scalar field leads to the formidable problem of finding
interesting solutions of the coupled field equations, though there have
been some promising attempts to find exact solutions of field
theoretical models that are even more complicated [PW]. A point particle
may be assumed to move on a geodesic of the field, but we know of no
theory of several interacting particles, or of continuous distribution
confined to a moving membrane, that satisfies the requirement of
invariance under general coordinate transformations. This real difficulty
can be avoided by postulating a physically reasonable matter
distribution, simple enough that Einstein's field equations can be
solved. The first attempt in this direction was that of Oppenheimer and
Snyder [OS]. Alternatively, but this really amounts to essentially the
same thing, one postulates an expression for the metric and studies the
implications for the distribution of matter. This can be a very
instructive exercise, as shown by the work of Einstein, Infeld and de
Hoffman on the motion of several particles [EI].

In the context of mass accretion of a black hole it is useful to
maintain rotational symmetry.   When we postulate a
metric that is Schwarzschild everywhere except on a moving sphere we
obtain a metric that depends only on the trajectory of the matter sphere.
The concentration of matter on an infinitely thin sphere is an
idealization that is not less physical than the idea of  point particles.
The trajectory depends on the interaction between the particles and it is
not difficult to believe that a smooth trajectory corresponds to
physically reasonable interactions.

The simplest example is the static case, $\dot f = 0$. The 
interpretation is especially simple, since the energy
momentum tensor is then diagonal. There is an energy density, radial
pressure and transverse pressure. The formulas obtained for the
Einstein tensor imply that the pressure (as defined in the manner first
suggested by Tolman [To]) is equal to the negative of the energy density.
This is unusual, but are not convinced that it is unphysical, especially
since one cannot be sure of what is physical when one is dealing with such
a highly idealized situation as is presented by an infinitely thin
spherical distribution of matter.  

If the internal forces do not balance the gravitational inward pull
then the matter sphere will move, inwards or outwards. If the repulsive
forces between the particles of the mass sphere are weak or absent we
expect a behaviour close to radial geodesic motion. For a test particle
with negligible mass in a Scwarzschild metric this would imply, at all
times, a relation
$$
\epsilon^2 \dot f^2 = (1-2m/r)^2 (\epsilon^2-1+2m/r),
$$
where $\epsilon > 0$ is a measure of the energy. The radial geodesics
approach the horizon as $t$ tends to infinity. In
our situation a reasonable approximation may be
$$
\epsilon ^2 \dot f^2 = (1-r_0/r)(1-r_1/r)(\epsilon^2 -1 + {r_0+r_1\over
2r}).
$$
This {\it ad hoc} formula represents a compromise between what looks like
geodesic motion from the inside/from the outside of the matter
distribution.  One may expect that within a wide family
of interparticle dynamics, there will be ingoing trajectories that
approach the horizon for large $t$.

The possibility that the in-falling matter may eventually penetrate the
horizon of its own creation seems to us to be extremely unlikely. If we
give the mass shell a little thickness we see that a particle on the
outside surface is subject  to attraction by the mass of the shell,
approximately equal to the attraction arising from a star in newtonian
gravity. If this attraction would be sustained, then the particle would
follow a Schwartzschild geodesic and then it would `never' reach the
horizon. In fact, the attraction diminishes as the particle moves through
the shell. The particle may be expected, in the absence of interactions
with the other particles in the shell,  to oscillate within the thickness
of the shell, and to fall inwards with a mean velocity that would be less
than that of the Schwarzschild null geodesic.

If this argument is accepted then there seems to be no reason to suppose
that loss of information is characteristic of physics in the
Schwartzschild metric. We have proposed a model of black hole creation
and growth in which the horizon is never approached, let alone penetrated.
 
It is certainly
interesting to study a family of reasonable physical models of matter
spheres in much greater depth. A statistical approach leads to
2-dimensional thermodynamics, with equations of state and a concept of
temperature. In some models the temperature will rise during the inward
motion and this would result in radiation. There is already an important
literature on this subject [TPM], [PW].

The Membrane Paradigm is a very interesting theory of black hole dynamics.
The horizon is treated as a physical membrane possessing  
characteristics similar to a soap bubble with negative surface tension
(positive pressure, as required to resist gravitational attraction).
In our opinion this is logically possible and not at variance with the 
correct interpretation of Schwarzschild solution, for the horizon is the
boundary of spacetime and Einstein's equations must be supplemented by
boundary conditions. The review by Damour [D] is full of ideas that can
be usefully applied within our framework, though it has more to
contribute to an understanding of rotating and charged black holes.  
However, the  
membrane paradigm is concerned with an already formed black hole; it does
not in itself  deal with the creation of a black hole, although it
stimulates certain ideas about black hole formation [H].  

 Parikh and  Wilchek  have proposed [PW] to move the membrane out from the
horizon, and give it physical reality, and to apply the ideas 
behind the mebrane paradigm to this new situation. Their proposal is quite
consistent with our ideas about mass accretion. It may be pointed out
that the presence of a membrane effectively hides the region where one
expects (when the outside metric is extended too far) to find the
horizon. So that Parikh and Wilczek, like ourselves,  deal with a
Schwarzschild metric but not with a Black Hole.
\b\b
\no{\it 3.3. Bianchi Identities and equations of motion.}

In the absence of singularities, the (contracted) Bianchi identities,
$$
{T^{\mu\nu}}{;\nu} = 0
$$
are true identities. However, in the presence of a point particle the
metric components become singular functions (distributions), such as
$1/r$, and meaningless
quantities such as $(1/r)\Delta{g_{00}} $ may appear in the field
equations. The discovery that these singularities, dipoles in Einstein's
terminology [EI], could be eliminated by postulating precise equations of
motion for the particles, was  a fundamental discovery that set  the 
theory of General Relativity apart from all other field theories.
In our context, with step functions
$\theta(r-f)$ in the expressions for the metric, one may encounter
meaningless quantities like $\theta(r-f)\delta(r-f)$. It occurred to us
that the trajectory of the mass sphere may be derived from the
requirement that such terms cancel out.

As we pointed out already in a footnote, the calculation of the Einstein
tensor does not encounter any untoward singularities. This is not
discouraging, for in the case of particles it was not a direct
examination of the Einstein tensor that led to equations of motion, but
rather the  Bianchi identities. Einstein insists on this.

The literature subsequent to the final summary provided by Einstein
and Infeld in 1950 [EI] is remarkably confused. The Bianchi identities are
just that, relations satisfied identically by the Einstein tensor.
So if the energy momentum tensor is identified with the tensor field that
satisfies the equation $G^{\mu\nu} = 8\pi T^{\mu\nu}$, then the relation
$$
{T^{\mu\nu}};\nu = 0
$$
is empty. And yet apparently not quite empty! We have attempted to
resolve this apparent contradiction as follows. Even if the components
of the Einstein tensor are well defined distributions,  delta 
functions in the context of point particles, the same need not be true of
the covariant derivative, notably the product of a component of $G$ by a
component of the connection, since the latter are not in the Schwarz
space of good functions. 

This suggests that the elimination of such
products should be deemed necessary, and that this is could lead   to 
equations of motion. However, extensive calculationsthat we
abstain from reporting lead to the conclusion that no such thing occurs in
our case.

In fact, this result is easily supported by an intuitive argument.
The result of Einsterin, Infeld and de Hoffman reveals a truth about
isolated systems. In order that a point particle deviate from geodesic
motion it must have an engine; this implies the emission of some kind of
exhaust that contributes to the energy momentum tensor and violates the
initial assumption that this quantity vanish away from the location of the
particle. The same conclusion surely applies to a compact and
isolated system: If each particle of a compact an isolated system moves
on a geodesic then there is no exhaust. However, absence of exhaust does
not imply geodesic motion, since the forces may be entirely internal
to the system. For example, a membrane may expand or contract due to the
existence of surface tension.

Consequently, we can accept the conclusion that no general principle
governs the trajectory of the incoming (or outgoing) spherical shower,
except for our strong conviction that no reasonable physical assumptions
will allow the shower to penetrate the horizon of its own making.

Quantization will alter the picture to some extent. The incoming (or
outgoing) massive shell will radiate gravitational waves and these will
produce particle pairs. It was suggested that this process can be
related to Hawkings proposal [H] concerning radiation from a black hole.
But to justify an exclusive preoccupation with this type of radiation
requires   strong assumptions: (1) that there are no 
`internal cosmic rays' (see  Section 2.2);\break  
 (2) that one has a
complete knowledge of the forces acting within the shell and (3) that the
radiation is described by a model quantum field theory. 
Thus Boulware [B] is able to reproduce Hawking's formula by assuming 
that there are no internal cosmic rays, no radiation resulting from
the interactions between the particles within the shell, and a model
of a scalar or spinor field. (This paper [B] considers a thin mass
shell but neglects the influence of the mass shell on the metric and this
allows to envisage that
 the collapsing object - the mass shell - may penetrate the
event horizon, which leads to a discussion of ``physics" inside the
horizon.)      
 \b
\ce{\steptwo Acknowledgements}
 
I thank Robert W.  Huff for very useful discussions.
 \b
\ce{   \bf REFERENCES}

[B] D.G. Boulware, Phys.Rev.D {\bf 13}, 2169 (1975).

[C] S. Chandrasekhar {\it The mathematical theory of black holes}, Oxford
Univ. Press, 1983.  

[E]~ A.S. Eddington, {\it The nature of the physical world}, Cambridge
University Press 1928. 

[EI] A. Einstein and  L. Infeld, Can.J.Math {\bf 1}; A.Einstein, L.
Infeld and B.Hoffman, 

\quad \quad Ann.Math. {\bf 39}, 65-301 (1938).

[F]~~ D. Finkelstein, Phys. Rev. {\bf 110} 965 (1959).

[Fr] ~C. Fronsdal Phys.Rev. {\bf 116} 778 (1959).

[H]~~ S.W. Hawking, Nature {\bf 248} 30 (1974). Preprint hep-th/0507171.

[K]~~ E. Kasner, Am.J.Math. {\bf 43} 126 and 130 (1921).

[Kr]~ M.D. Kruskal, Phys.Rev. {\bf 119} 1743 9(1960).

[OS] ~J. R. Oppenheimer and H. Snyder, Phys. Rev.
56, 455 (1939)

[PW] M.K. Parich and F. Wilzcek, arXiv:gr-qc/9807031v2 4Feb 2000. 

[S]~~~~ K. Schwartzschild, Sitz.Preuss.Akad.Wiss. 1916, 189.

[S]~~~~ G. Szekeres, Publ.Mat.Debrecen {\bf 7}, 285-301 (1960).

[TPM]	K.S. Thorne,  R.H. Price and  D.A. Macdonald,  

\r  {{\it The
membrane paradigm}, 
Yale Univ. Press, 1986.}  

[T] R.C. Tolman,
{{\it Relativity, thermodynamics and
cosmology,}  Clarendon press, 1934.}
\end

\end

\ce{ \Blue \bf APPENDIX} 
The components of the connection, copied from Tolman or calculated
afresh, are
$$
\matrix{\hskip1mm \G_{11}^1 = -\nu'/2, &~~ \G_{11}^4 = -e^{-2\nu}\dot
\nu/2, &
 \G_{12}^2 = \G_{13}^3 = 1/r,\cr \G_{14}^1 = -\dot \nu/2, &
\hskip-6mm\G_{14}^4 =
\nu'/2, & 
 \hskip-6.8mm\G_{22}^1 = -re^\nu, \cr \hskip-.8mm\G_{23}^3 = \cot\theta, &
\hskip5.9mm\G_{33}^1 = -r\sin^2\theta~e^\nu, & \hskip4mm\G_{33}^2 = - \sin
\theta\cos\theta,
 \cr   
\hskip3.9mm\G_{44}^1 = e^{2\nu}\nu'/2, &  \hskip-6mm\G_{44}^4 = \dot
\nu/2.
\cr}
$$

We calculate,
$$\eqalign{&
 {T_4^\nu}_{;\nu} = {T_4^\nu}_{,\nu} - \G_{4\nu}^\lambda T_\lambda^\nu +
\Gamma_{\lambda\nu}^\nu T_4^\lambda
\cr}
$$
Writing this out in detail, but omitting 4 terms that cancel identically,
for any connection, we have
$$
=  {T_4^\nu}_{,\nu} - \Gamma_{41}^1T_1^1 - \G_{44}^1T_1^4 + (\G_{11}^1 +
\G_{12}^2 + \G_{13}^3)T_4^1 + \G_{41}^1T_4^4.
$$
Two terms cancel because $T_4^4 = T_1^1$ and there remains
$$\eqalign{&
=  {T_4^\nu}_{,\nu}   - \G_{44}^1T_1^4 + (\G_{11}^1 +
\G_{12}^2 + \G_{13}^3)T_4^1\cr
& = {T_4^\nu}_{,\nu}  -(\nu'/2) e^{2\nu}T_1^4 + (-\nu'/2 + 2/r)T_4^1.
\cr}
$$
There is a hint of trouble here, although a  calculation {\it ab initio} 
gives
$ 
e^{2\nu}T_1^4 =   -T_4^1,
$ 
so that we end up with the simple and correct result
$$
{T_4^\nu}_{;\nu} = {T_4^\nu}_{,\nu} + (2/r)T_4^1 = 0.
$$
Though one may have some difficulty in swallowing the last step, one
hardly sees any hope for relating this difficulty to the equations of
motion.

Next, 
$$\eqalign{&
 {T_1^\nu}_{;\nu} = {T_1^\nu}_{,\nu} - \G_{1\nu}^\lambda T_\lambda^\nu +
\Gamma_{\lambda\nu}^\nu T_1^\lambda
\cr}
$$
Omitting four terms that cancel we get
$$\eqalign{&
  = {T_1^\nu}_{,\nu} - \G_{11}^4 T_4^1 - 2\G_{12}^2
T_2^2 - \G_{14}^4T_4^4 + 2\G_{12}^2T_1^1 +  \G_{41}^4T_1^1 +
\G_{44}^4T_1^4
\cr}
$$
Two terms cancel because $T_4^4 = T_1^1$ and there remains
$$
= {T_1^\nu}_{,\nu} + (\dot \nu/2)e^{-2\nu}T_4^1 -(2/r)T_2^2   
+(2/r)T_4^4 + (\dot \nu/2) T_1^4.
$$
here too, if we are permitted to take into account the relation $ 
e^{2\nu}T_1^4 =   -T_4^1,
$ 
we are led to a simple and correct result that
$$
 {T_1^\nu}_{;\nu} = {T_1^\nu}_{,\nu} -(1/r)(T_2^2 + T_3^3) + (2/r)T_4^4 =
0.
$$ 

The needed relation, $ 
e^{2\nu}T_1^4 =   -T_4^1,
$ 
expresses the symmetry of the contravariant energy momentum tensor.
To preserve this symmetry we should not be dealing with the mixed tensor.
So let us examine, as does Einstein,
$$\eqalign{
{T^{4\nu}}_{;\nu} - {T^{4\nu}}_{,\nu}  =& ~\G_{\nu\lambda}^4 T^{
\lambda\nu} + \Gamma_{\nu\lambda}^\nu T^{4\lambda}\cr
  = &~\G_{11}^4T^{11} + 2\G_{14}^4T^{41} + \G_{44}^4T^{44}\cr
& ~~~+ 
(\G_{11}^1 + \G_{12}^2 + \G_{13}^3 + \G_{14}^4)T^{14} + (\G_{41}^1 +
\G_{44}^4)T^{44}\cr
=&  -(\dot \nu/2)e^{-2\nu}T^{11}  + (\dot\nu/2) T^{44}
+(\nu' + 2/r)T^{14}.
\cr}
$$
The $\G$-terms can be combined as in the other cases so as to write this
as follows,
$$\eqalign{
{T^{4\nu}}_{;\nu} = {T^{4\nu}}_{,\nu}   +\Big[{r_1-r_0\over  
A\ln{r-r_1\over r-r_0}}  + {1\over r} 
 }\Big]T^{14}.
$$
Finally,
$$\eqalign{
{T^{1\nu}}_{;\nu} - {T^{1\nu}}_{,\nu}  =&~ \G_{\nu\lambda}^1 T^{
\lambda\nu} + \Gamma_{\nu\lambda}^\nu T^{1\lambda}\cr
  = &~\G_{11}^1T^{11} + 2\G_{14}^1T^{41}  +
\G_{22}^1T^{22} + \G_{33}^1T^{33} + \G_{44}^1T^{44}\cr & ~~~~~~+ 
(\G_{11}^1 + \G_{12}^2 + \G_{13}^3 + \G_{14}^4)T^{11} + (\G_{41}^1 +
\G_{44}^4)T^{14}\cr
=& ~(-\nu'/2 + 2/r)T^{11} -r e^\nu(T^{22} + \sin^2\theta\, T^{33})
- \dot\nu T^{14} + e^{2\nu}( \nu'/2) T^{44}\cr
 =&~ e^{2\nu}(\nu'/r^2) -e^{2\nu}(2/r^3) -
e^\nu(\nu'/r^2) + e^\nu(2/r^3) 
     + (1/r) \ddot\nu + (1/ 2r)(e^{2\nu})''\cr
=&~ (2/r)T^{11}  + P\big[(1/r) \ddot\nu + (1/ 2r)(e^{2\nu})''\big], 
\cr}
$$
where the symbol $P$ stands for an uncompensated remainder, namely 
$$\eqalign{
P\big[(1/r)  \ddot\nu + (1/ 2r)(e^{2\nu})''\big]& 
  := {1\over r}\ln{r-r_1\over r-r_0}\Big(\dot f^2\delta'(r-f) - \ddot
f\delta(r-f)\Big)\cr
&~~~~~~~~~~~~ -{\Delta r\over  r^2}(1 -{r_1+r_0\over 2r})\delta'(r-f),
\cr}
$$
with $\Delta r = r_1-r_2$. This contains second order derivatives and it
cannot, in general, be expressed in terms of the components of the energy
momentum tensor.
  The only way that this expression can be reduced is if
there are functions $\alpha_{1,2,3}$ of $r$ such that 
$$
P\big[(1/r)  \ddot\nu + (1/ 2r)(e^{2\nu})''\big] = \alpha_1(r)T^{22}
+ \alpha_2(r)T^{11} + \alpha_3(r)T^{44}.
$$
This imposes no conditions on the trajectory.
\ve

Explicitly,  
$$\eqalign{&
{1\over r}\ln{r-r_1\over r-r_0}\Big(\dot f^2\delta'(r-f) - \ddot
f\delta(r-f)\Big)  -{\Delta r\over  r^2}(1 -{r_1+r_0\over
2r})\delta'(r-f)\cr
&  = \alpha_1(r)\Bigg({-\Delta r\over
2r^3}\Big( 1+ {\dot f^2\over 
 A}\Big) \delta'(r-f) 
   + {\Delta r\over 2r^3A} \ddot f 
\delta(r-f)\Bigg)\cr
&~~~~~~~~~~~~~~~~~~~+\alpha_2(r){\Delta r \over
r^2} \Big( 1-{r_1 + r_0\over 2r}\delta(r-f) 
\Big)+ \alpha_3(r) {1\over r}\ln{r-r_1\over r-r_0}\delta(r-f) 
\cr}. 
$$
This amounts to two conditions,
$$
\alpha(r)(1 + \dot f^2/A) = -{2r^2\over \Delta r}\ln{r-r_1\over
r-r_0}\dot f^2 + 2r(1 -{r_1+r_0\over 2r}).
$$
and 
$$  
-\Bigg({1\over r} \ln{r-r_1\over r-r_0}+    
    \alpha_1(r){\Delta r\over 2r^3A}\Bigg) \ddot f =\alpha_2(r){\Delta r
\over r^2} \Big( 1-{r_1 + r_0\over 2r} 
\Big)+ \alpha_3(r) {1\over r}\ln{r-r_1\over r-r_0},
$$
The crux of the matter is the compensation of the terms involving the
distribution $\delta'(r-f)$,
a differential equation that determines the trajectory up to a constant
parameter,
$$
\dot f^2 = A{2r-r_1 - r_0 -\alpha(r)\over 
\alpha_1 +   2r^2A \ln{r-r_1\over r-r_0}/\Delta r}.
$$
If $\dot f^2$ is to be positive for large $r$ then we must have 
$$
 \alpha_1 = 2r -r_0-r_1 + \beta(r)/r,
$$ 
with $a(r)$ constant mod$(1/r)$; so that 
$$
\dot f^2 = A{  \beta(r)/r\over 
-\beta  -2r+r_0+r_1 -   2r^2 A  \ln{r-r_1\over r-r_0}/\Delta r}.
$$
   To obtain $\dot f^2$ to the two leading orders in $1/r$ we must expand
$$\eqalign{&
1/A = 1 + {r_1 + r_0\over r} + {r_1^2 + r_1r_0 +  r_0^2\over r^2}   
+{r_1^3 + r_1^2r_0 + r_1r_0^2 + r_0^3\over r^3} + ...,\cr
& r\ln {r-r_1\over r-r_0} = -\Delta r(1 + {r_1 + r_2\over 2r} +
{r_1^2 + r_1r_0 +  r_0^2\over 3r^2} + {r_1^3 + r_1^2r_0 + r_1r_0^2 +
r_0^3\over 4r^3} + ...). 
\cr}
$$
To two leading orders in $1/r$ it is
$$
\dot f^2 = (1-{r_1 + r_2\over 2r}){\beta(r)\over (1/3)(\Delta r)^2(1+{r_0
+ r_1\over 2r})  - \beta(r)}, 
$$

\ve
\ce{\steptwo\Blue Newtonian Model}\Black

Consider a spherical shell with inner diameter $r_0$, outer diameter
$r_1$ and uniform density (Cartesian coordinates).
The Newtonian potential of a volume of unit density at a distance $r$
from the origin is
$$
U(r) = -{4\pi\over 3}r(r^3 - r_0^3) - 2\pi(r_1^2 - r^2).
$$
The total potential of the shell is
$$
V  =  {(4\pi)^2\over 30}(r_1-r_0)^2(2r_1^3 + 4r_1^2r_0 + 6r_1r_0^2
+ 3r_0^2),
$$
or to lowest order in $(\Delta r)^2$,
$$
V(R,\Delta r) ={M^2\over 2R},~~ R = {r_1 + r_0\over 2},~~ M = 
4\pi R^2\Delta r. 
$$
The kinetic energy is, in the same approximation, is $ M\dot R^2/2$, and
the equation of motion can integrated to
$$
M\dot R^2 - M^2/2R = 2E = {\rm constant}.
$$
This is on the assumption that the total mass, not the thickness of the
shell, is constant.

\ce{\steptwo References}

[Ei] Albert Einstein, Canadian J. Math. volume 1 1950.

To lowest order in $1/r$ we must have $\alpha(r) = 2r$, so we set
$$
1/\alpha(r) = 1/2r - a/2r^2 + o(1/r^3),
$$
with $a$ a constant parameter. Then to relative order $1/r^2$
$$
\ln{r-r_1\over r-r_0} = -{\Delta r\over r}(1 + {r_1 + r_0\over 2r})
$$
and to first order in $1/r$ our equation reduces to
$$
\dot f^2(r_1 + r_0)  = \dot f^2(a -{r_1 + r_0\over 2}) +a -
{r_1+r_0\over 2},
$$
or
$$
\dot f^2 = -{a- (1/4)(r_1 + r_0) \over a -  (3/4)(r_1 + r_0)}  +
o(1/r).
$$
This must be positive, so
$$
(1/2)(r_1 + r_0) < a < (3/2)(r_1 + r_0).
$$ 
To obtain the equation of motion to the next order set
$$
1/\alpha(r) =  1/2r - a/2r^2 - b/r^3 + o(1/r^4),
$$

\ve

\no{\it 2.4. Several mass spheres.}

The creation of a black hole is an idea that contains an internal
contradiction. We can envisage, nevertheless, a situation in which the
metric is piecewise Schwarzshild, and undergoes sporadic changes that, to
a local observer, amounts to a sudden increase in the mass. 

Imagine a series of mass spheres at $r = f_1,...,f_n$,
and the metric
$$
ds^2|_f = \Bigg\{\matrix{\hskip-9mmds^2(r_0),~~~ r_0<r<f_1(t),\cr
\hskip-.5in...~~~~,\hskip.5in...\cr 
 ds^2(r_k),~~~ f_k(t)<r<f_{k+1}(t),\cr
\hskip-.5in~...~~~~,\hskip.5in...\cr \hskip-1.9cm ds^2(r_n),~~~f_n <
r.}\eqno(2.4) 
$$
 with parameters $r_0 < ...<r_n$ and $r_k<f_k(t),~ k =
1,...,n$. During some time interval, when $r_n << f_1(t)$, an observer 
at $r_n < r < f_1(t)$ sees a Schwarzschild metric with mass parameter
$m_0 = r_0/2$. If now the first mass sphere moves inwards  past our
observer, then,  if he is not traumatized by this,
he will notice that the mass of his black hole has increased; that the
horizon has moved outwards. Each time that an inward moving mass sphere
goes past, he sees an  increase in the Schwarzshild mass. If $r_0 = 0$ he
observes the creation and subsequent growth of a black hole out of a local
Minkowski space; this is perhaps the most interesting
possiblity. Eventually he may find himself squeezed between
$r_k$ and $f_k(t)$, but as it takes an infinite time for $f_k(t)$ to
reach $r_k$ there is time to escape.

\b

\no{\it 2.3. Speculation about the nature of the incoming
matter.}

It has occurred to us to study the energy momentum tensor in terms of
co-moving coordinates. According to Tolman [To] as well as Oppenheimer
and Snyder [OS] these are coordinates in which the components $T_1^4$ and
$T_4^1$ of the energy momentum tensor vanish. As far as we have been able
to discover, no one comments on the fact that there is no coordinate
transformation of the type $t,r \rightarrow T = T(t,r),~R = R(t,r)$
that would accomplish this for the tensor (2.5).

Let us go further and imagine that the mass spheres are so densely packed
as to approach a continuous distribution of mass, with mass parameter
 $r_k/2,~ 0<k<1$ and $f_k(t) =  r_k/p(t),~ p(t)<1$. The metric is then
$ ds^2(r_0),~ r_0 < r < p(t)r_1$,~  $ ds^2(r_n),~ p(t)r_n< r$. Between
these limits it is 
$$
\Big(1- p(t)\Big)dt^2 - \Big(1- p(t) \Big)^{-1} dr^2 - r^2d\Omega^2.
$$ 
The energy momentum tensor is non-zero in this interval only, with
components 
$$\eqalign{&
8\pi{T_1}^1 =  8\pi{T_4}^4 = 0,\cr
& 8\pi{T_2}^2 = 8\pi{T_3}^3 = ...\cr
&
8\pi{T_4}^1 = ..., ~~8\pi_{T_1}^4 = ...
\cr}
$$  
$$\eqalign{&
8\pi
\sqrt{-g} T^{11} = {1\over 2r} \Big(e^{k/2}(1-2m_0/r)^2-(1-2m_1/r)^2 
\Big)\delta(r-f),\cr
& 8\pi \sqrt{-g}T^{44} =  {1\over r}\Big(
e^{k/2}\ln(1-2m_1/r) - e^{-k/2}\ln(1-2m_0/r)\Big)\delta(r-f),\cr 
  & 8\pi\sqrt{-g} T^{41} = 8\pi\sqrt{-g} T^{14} =   8\pi
\sqrt{-g}T^{44}\dot f,\cr
& 8\pi8\pi \sqrt{-g}T^{22} = {1\over r^3}(e^{k/2}-1)\delta(r-f) + {1\over
2r^2}\Big(e^{k/2} -1+ {2\over r^2}(m_1 - m_0e^{k/2})\Big)\delta'(r-f)\cr
& +{1\over 2r^2}(1-2m_0/r)^{-1}(1-2m_1/r)^{-1}\Big( e^{-k/2} -1 + {2\over
r^2}(m_0 - m_1e^{-k/2})\Big)\dot f^2\delta'(r-f)\cr
& +{1\over 2r^2}(1-2m_0/r)^{-1}(1-2m_1/r)^{-1}\Big(e^{-k/2} -1 + {2\over
r^2}(m_0 - m_1e^{-k/2})\Big)\ddot f\delta(r-f).
\cr}
$$